# Assessing the impacts of convening experts: a bibliometric analysis of a research program spanning four decades

Deborah M. Buehler[1*], Mark J Daley[2], and Kyle Demes[3]

[1]*Research Analytics, Office of the Vice President Research and Innovation, University of Toronto, Toronto, Ontario, Canada. 0000-0003-3669-6364 *d.buehler@utoronto.ca*

[2]*Department of Computer Science, Western University, London, Ontario, Canada / The Vector Institute for Artificial Intelligence, Toronto, Canada. 0000-0002-6939-9772*

[3]*Demes Strategic Consulting, Port Moody, British Columbia, Canada. 0000-0003-2780-0393*

Running head: Assessing the impacts of convening experts

## Abstract

Over the last few decades, research institutions and funders have begun policies and programs that incentivize large-scale collaboration across institutions on focal research areas. Yet, few studies have evaluated the impact of those programs on research, particularly on timelines longer than a few years. Using the Canadian Institute for Advanced Research (CIFAR) as a case study, we examined the impacts of supporting a research program that convened experts across intuitions and countries for 40+ years. In this study, we used the Scopus bibliometric database to analyse publishing and citation trends within this team since its formation in 1986 and used nearest neighbour matching to compare these trends against authors across the globe with similar career characteristics to measure how effectively the CIFAR program Gravity & the Extreme Universe (CIFAR-GEU) has catalyzed collaborations and produced high quality research outputs. We found a greater degree of co-authorship within the CIFAR-GEU group compared to the Control group. We also found that the outputs generated by the CIFAR-GEU group had, overall, higher values for citation-based impact indicators (e.g., stronger metrics around citations, impact from international collaborations and reach beyond academia).

## Introduction

Our world and its societies face complex challenges and tackling these "wicked problems" will require that the world's best minds work together in teams with other experts with diverse perspectives (Rittel and Webber 1973). To address these needs, research institutions and funders around the world have created new funding programs specifically targeted at creating collaborations among leading researchers in a topic. These have generally taken one of two forms: awarding teams or institutions large grants to fund research directly, or giving teams smaller, strategic funds to convene and collaborate on ideas and knowledge creation, but not directly funding research.

Examples of the former include the Canada First Research Excellence Program (CFREF) and Horizon 2020 in Europe. CFREF was established in 2014 and funds large-scale strategic research initiatives led by institutions (rather than individual researchers or projects) with funding in the hundreds of millions per grant (Government of Canada 2021). Similarly, Horizon 2020 was launched in 2014 by the European Union (EU) and funded research and innovation with companies both within and outside of the EU with a budget of nearly €80 billion (Mitra & Niakaros 2023). An example of the latter is the National Center for Ecological Analysis and Synthesis (NCEAS), established in 1995 by NSF to synthesize accumulating ecological information. Funding provides time, resources, and a creative environment for visitors to immerse themselves in collaborative synthesis (Hampton and Parker 2011). Another example of the latter approach is Canadian Institute for Advanced Research (CIFAR), a research organization based in Canada and established in 1982 that has been convening some the world's best minds to tackle complex questions for more than 40 years (Brown, 2007). CIFAR organizes meetings between interdisciplinary program members and other national and international guests. Funding covers the costs of planning, venue, travel, accommodation, and small grants towards collaboration in high-risk research areas related to the program's overarching goals. CIFAR's programs are funded on decadal timescales allowing research networks the time to grow and solidify.

The trend towards tackling complex problems in interdisciplinary groups is still relatively new, and few programs have existed long enough to demonstrate impact over the long term. There is also the question of how to measure such impact. Though there has been good progress developing methods for assessing the societal impacts of research, especially in the health field (Greenhalgh et al. 2016), contemporary concepts and methods of research impact assessment are not well suited for research that crosses disciplinary boundaries and intervenes in complex systems (Belcher and Hughes, 2021). Because CIFAR has been convening researchers for decades, a few of its programs provide a model for measuring connections created and impact generated over the long term. Furthermore, CIFAR has established processes involving expert review panels and multiple stages of evaluation to assess existing programs for renewal and to select new programs through a competitive global call.

In this study, we measured the impact of one of CIFAR's longest standing programs, the Gravity & the Extreme Universe program (hereafter CIFAR-GEU and formally known as Cosmology and Gravity). The program began in 1985 to study the origin of the universe and the astronomical structures seen within it. A few of the program's founders were already in Canada and had worked together, but they proposed bringing in experts from outside Canada, and bringing together knowledge from the then disparate fields of physical cosmology, particle physics and gravitational theory. In the early years, experts moved to Canadian universities from places such as Princeton, Stanford, and Cambridge. Later researchers worked together without moving as CIFAR became more global.

We chose this program because the topic lends itself to publication in traditional journals making co-publication data mined from an online bibliometrics database a useful proxy for both research outputs and collaborations generated. However, as with many fields, astronomy and cosmology has changed greatly over the 40 years and collaboration, co-authorship, and research output have increased for all researchers. Is it possible to extract CIFAR's impact from this background of changes?

Finding a control group or counterfactual is a key challenge in long-term impact studies; in this case, we are aware of no other network that is comparable to the cohort brought together in the GEU program over the past 40 years. We therefore used a novel method to create a control group to measure impact. This method involved mining author profiles through APIs in the online bibliometric database Scopus to create a counterfactual. We pulled all publications from all researchers involved in the CIFAR-GEU program over its nearly 40-year existence (n = 16,808). We then pulled data on all researchers (n = 26,177) who have co-authored with our CIFAR researchers throughout their careers, leveraging the high degree of co-authorship in this field as a way of identifying potential comparators. We then used nearest neighbour matching to pair each CIFAR-GEU researcher to a co-author from the larger pool with the most similar date of first publication (to control for career stage), total number of scholarly outputs over their entire career (to control for productivity), and the number of unique coauthors throughout career (to control for collaborative style), in that order.

Our method generated a Control group sampled from a pool of CIFAR-GEU co-authors and paired with GEU researchers while controlling for career stage, scholarly output and collaboration style. We reason that this control group represents a random—but CIFAR-level-caliber—sample of people working in the same field. To measure the impact of convening researchers on collaborative networks, we constructed co-authorship networks for the CIFAR-GEU and Control groups and compared them over time. In this way, the control group acts as a proxy for changes in the field overall. We predicted stronger relationships created and maintained over time in the CIFAR-GEU group compared to the Control group. Specifically, we expected that co-authorship density (the number of edges between nodes, where each node is an author) would increase in both groups because scholarly output and collaborations are increasing generally but hypothesized a much more pronounced effect in the CIFAR-GEU group. We also compared these two groups across several publication and citation metrics to study the impact of stronger and longer-term networks. We hypothesize that CIFAR-GEU authors have higher citations metrics, more impactful international collaborations and greater uptake of research outside of academia than the Control group.

## Methods

### Choosing the program and the proxy for collaboration

We chose the Gravity & the Extreme Universe program because it is one of CIFAR's longest running programs and because the topic lends itself to publication in traditional journals making co-publication data mined from an online bibliometrics database (Scopus/SciVal) a useful proxy for collaboration. We acknowledge that traditional publications are not the primary research output in all fields. In such cases, other, more appropriate data (e.g., co-supervised students, co-PI status on grants, or co-produced exhibits or performances) could be used as a proxy for collaboration.

### Data source and method for mining the online databases

We used the Scopus and SciVal bibliometric databases for all analyses in this study and used a combination of .com and API interfaces to extract data. Although some studies have identified limited coverage of research in social sciences and humanities in Scopus, it is generally considered to have comprehensive coverage of research in the physical sciences (Mongeon and Paul-Hus, 2016). The data required for this study could not be easily extracted from the web user interface and so we used API-based data extractions using the Elsevier Data Fetcher version 7.3.4 (Elsevier, 2023). Data analyses were performed in R Statistical Software Version 4.2.2.

We used CIFAR records to identify the 93 researchers who had participated in the GEU program from its inception to May 2022, the cut off for an upcoming program review after which the program began the process of "sunsetting". We then looked up their Scopus Author IDs and various bibliometrics in the Scopus database. Scopus profiles are algorithmically generated and not curated by the researchers themselves. Thus, they are fairly consistent globally, whether or not researchers and institutions use Scopus services. Elsevier reports author profile precision > 99% and recall on average 95% for author profiles (Elsevier 2019). We checked the profiles of CIFAR-GEU authors to ensure the Scopus ID matched the correct author and we submitted a revision for one CIFAR author whose profile was conflated with a researcher in an unrelated field. We similarly checked that publications broadly matched the correct field of study for matching authors in downstream analyses of the Control group (created below). However, we did not undergo a systematic clean-up at the individual publication level for either group.

### Creating a control group or counterfactual

Our goal was to create a control group comparable to, and of a comparable size, to our group of CIFAR-GEU researchers over the programs 40-year existence. To do this we used nearest neighbour matching of propensity scores to pair each CIFAR-GEU member with a coauthor having a comparable career stage (date of first publication), productivity (scholarly output), and collaborative style (the number of unique coauthors).

### Determining the pool of co-authors from which to sample

We pulled complete publication histories of all 93 CIFAR-GEU researchers on December 17, 2022. This dataset included standard bibliometric metadata on 16,808 publications and a full list of unique Scopus Author IDs for all 26,177 authors on those publications. We then pulled author-level metadata from Scopus Author Profiles for all Author IDs to be used in downstream analyses, including: scholarly output (i.e., count of documents indexed in Scopus), co-author count (i.e., total number of unique co-authors on Scopus-indexed documents), year of first publication (i.e., publication year of first Scopus-indexed document), and total number of citations (i.e., the total number of Scopus-indexed documents citing Scopus-indexed work by each other). Pulling data on all co-authors allowed us the greatest flexibility for null group pairing (e.g., determining how best to match the GEU fellows with a coauthor).

A common stumbling block when doing bibliometrics in physical science fields is that astronomers and physicists sometimes co-publish papers with >1000 authors. We wondered whether this might falsely inflate the pool of co-authors, and we checked whether there were just a few outlying papers that we could exclude or whether such multi-author papers were common. We found several thousand papers in our dataset with > 1,000 authors and most of the high author papers had 2,000 to 2,500 authors, indicating that large-scale collaborations are representative of the field. Furthermore, the hyper-collaboration among researchers on some publications helped ensure that the co-author pool was as comprehensive as possible for determining a control group. Therefore, we did not exclude any publications from analyses.

### Determining which career parameters to use for nearest neighbour matching

In our initial exploration, density plots comparing CIFAR's GEU researchers (n=93) with the entire pool of co-authors (N>26,000 unique co-authors) showed that CIFAR researchers generally began their careers earlier (i.e., have an earlier date of first publication), have higher publication volume than the global pool, and tend to be involved in more large-scale collaborations (Figure 1). These results make sense because, in the early years of the CIFAR-GEU program, most of the researchers selected for membership were already established researchers. We also noted a more pronounced bimodal distribution in collaboration style (total number of co-authors over career to date) for CIFAR-GEU members. Taken together, this indicates that there are distinctive productivity, career, and collaboration traits between CIFAR researchers and the general pool of authors.

Noting these trends and noting that the density plots show ample areas of overlap between the distributions for pairing (Figure 1, top row), we employed one-to-one nearest neighbour matching based on the full co-author pool, controlling for career stage, productivity style, and collaborative style (prioritizing variables in that order) to produce CIFAR-GEU and Control groups of equal size and comparable career stage and caliber. We performed the nearest neighbour matching of propensity scores using the R software package MatchIt v 4.5.0 (Ho and Imai, 2018; Köhler, 2016) to sample from within the larger pool of co-authors and to create a Control group of equal size to our group of CIFAR-GEU program researchers. We prioritized author matching by publication count, year of first publication, and number of unique co-authors, in that order, and were able to match all 93 CIFAR researchers to researchers in the co-author pool. Density plots comparing the CIFAR researchers with their matched controls (Figure 1 bottom row, both of sample size 93) show strong pairwise matching.

Restricting our control group to authors from within the larger co-author distribution and matching control authors to CIFAR-GEU researchers in terms of career stage, productivity, and collaboration traits allows us to compare research quality metrics between the two groups. We argue these pairings enable us to detect differences between the groups in research citation metrics while constraining differences.

## Testing for differences between the CIFAR and control groups

### Co-author networks

CIFAR brings together people who wouldn't otherwise collaborate to generate ideas from a synergy that wouldn't otherwise happen. To visualize and measure our success at doing this, we generated co-author networks and compared them between the CIFAR-GEU and Control groups.

We extracted author networks for both Control and CIFAR-GEU authors from Scopus on March 1, 2023 using the Data Fetcher in an adjacency list format with nodes represented as Scopus Author IDs and edges represented by Scopus Publication EIDs for each node-to-node combination. Publication EID was used to pull year of publication metadata from Scopus, facilitating the comparison of networks over time. The author network adjacency lists were ingested into R and we used igraph (Csardi 2010) to analyse and visualise networks.

We chose to arrange the nodes in the network in a circular rather than the traditional ball and stick layout, because standardizing node placement in a circle with the edges spanning between, made comparing the density of edges across many nodes more easily comparable between the two groups over time.

### Comparing metrics using group- vs. author-level analyses

When comparing bibliometric indicators across two groups of authors, two distinct approaches are common: (1) **author-level analyses** which compute metrics for each author in each group by aggregating the metrics of the publications by each author individually, facilitating comparisons at the author level. (2) **group-level analyses** which pool all unique publications from any of the authors into a single publication set to facilitate direct comparison of the publication sets produced by the groups. The former can be helpful when comparing the researchers in the two groups and the latter can be helpful when evaluating differences in the work produced by the groups.

Evaluating researchers at the author-level is a common practice. For example, when a researcher goes for tenure, the publications they author are counted (in some disciplines taking into account author order) even though the same publications (or a subset) can also be used to evaluate a co-author in a different tenure review. Similarly, researchers are the foundational unit in CIFAR program creation and award selection, thus author-level analyses are of critical interest in program evaluation.

However, especially when looking at citation bibliometrics, high citation metrics in a few publications co-authored by many CIFAR-GEU authors could be responsible for amplifying differences between groups at the author-level. For instance, the number of publications shared by co-authors likely differs between the CIFAR-GEU and Control groups (see network analyses below). In the CIFAR-GEU group many of publications are likely to be co-authored with other members of the group and so would be used in the determination of author-level metrics for multiple authors. In the Control group, we expect fewer collaborations amongst the selected researchers. Furthermore, because we built the control group from co-authors of CIFAR-GEU researchers, by definition, there will be publications shared between the CIFAR-GEU and Control groups. Indeed, we found that 2055 publications of the about 13,500 (15%) examined between 1996 and 2021 were shared between the CIFAR-GEU and Control group.

For the reasons above, we present analyses at both the author- and group-levels. Data for the author-level networks and density plots are from Scopus. Scopus allows analyses of complete, current publication histories, and we chose to examine author-level data from pre-1987 (pre-CIFAR) to Mar 2022. Data for the group-level comparison between the CIFAR-GEU and Control groups is taken from SciVal. SciVal can compute groups for which any publication authored by more than one group member is counted only once. However, the SciVal database has time constraints, thus for the group-level bibliometric comparisons data are from 1996 (the first available year) to 2021 (the last complete calendar year).

We chose a panel of bibliometric variables to compare the two groups. These metrics measure aspects of the quality of scholarly output (mainly citation metrics), international collaboration, and reach beyond academia (Table 1). We chose these metrics because many of them are snowball metrics, which means that they are tested, agreed upon and methodologies (Colledge, 2017). They also measure aspects of the bibliometrics universe that we find the most important in terms of measure the success of a CIFAR program (impact, collaboration and reach). Note that we do not compare Scholarly output, because this is one of the metrics we used for the pairwise matching and thus we know that this metric will be similar across the groups.

## Results

### Co-author networks

CIFAR convenes researchers to build and strengthen networks. Figure 2 shows co-author networks for both groups at three timepoints, publications pre-1987 before the CIFAR-GEU program was started, mid-way, and as of March 2022. In this analysis, the Control group does not represent an existing team, rather represents baseline collaborations in the field globally. Still, the difference between the grey and the red networks is striking. Both groups have increased collaboration (co-authorships) over time, but with the CIFAR-GEU group, the cumulative effect is that by March 2022, the circle is filled with edges (each edge is a co-authored publication) because everyone within the CIFAR program is collaborating and co-publishing.

### Group- and author-level comparisons of CIFAR-GEU and Control groups

Figure 3 shows group- and author-level analyses for three citation metrics. The line graphs on the left show group-level data comparing metrics on publication sets for each group over time from 1996 to 2021. The density plots on the right show author-level data for the whole period from 1986 to 2021.

The line graphs show that CIFAR-GEU authors' publications have more citations, more citations per publication, higher FWCIs, and that this difference persists over time. Number of citations and citations per publication both trend towards lower numbers in more recent years because it takes years for publications to accumulate the bulk of their citations. However, the trend of higher citations for CIFAR-GEU groups persists even in the most recent year. Field-Weighted Citation Impact normalizes across year of publication (along with research area and document type) when determining citation impact and so the difference in recent years is more pronounced by this indicator.

The density plots showing the distribution of author-level metrics show the same trend for each of the citation indicators. For all three of the citation metrics, the median is shifted to higher values for the CIFAR-GEU authors and the density distributions for the CIFAR-GEU authors are flatter, more right-weighted, with a substantially longer right tail. The consistent trends at the group- and author-levels suggest that results are not an artifact of multi-counting publications, especially in the CIFAR-GEU group.

Table 2 shows further comparisons between publications sets from the CIFAR-GEU and Control groups across metrics measuring the quality of output, international collaboration, and impact, and reach beyond academia.

Group level-publication sets show that the CIFAR-GEU group excels in measures of publication quality. CIFAR-GEU publications are cited more, both in terms of raw citations and citations standardized by overall output (citations per publication). A slightly higher percentage of the CIFAR-GEU group's publications have been cited than the Control group and a higher number of them are in the top 1% of highly cited works. Finally, Field-weighted Citation Impact (FWCI), is higher in the CIFAR-GEU group. A FWCI above 1 indicates that publications in the set are cited more than the average for publications in that field, year of publication, and document type. Both groups are above average, which is not surprising because we generated the Control group from CIFAR-GEU co-authors matched for similar career stage, productivity style, and collaborative style.

In terms of international collaboration, the percentage of international collaboration is slightly higher in the Control group. However, the CIFAR-GEU group has a higher impact from their publications with international co-authorship.

Finally, in terms of reach beyond academia, the CIFAR-GEU group had slightly lower percentage of outputs co-authored by researchers from both academic and corporate or industrial, affiliations; however, like with international collaboration, the CIFAR-GEU group had higher impact with the academic-corporate collaborations they had. Additionally, the CIFAR-GEU group had a lower number of patents citing their scholarly, but a higher number of policy documents referencing their outputs, as well as blog mentions, news mentions and tweets.

## Discussion

CIFAR convenes groups to create synergies that generate unique research. Our aim for this study was to test whether CIFAR brings together researchers who wouldn't otherwise have collaborated and that the networks CIFAR creates have an impact. To do this we generated a Control group sampled from a pool of CIFAR-GEU co-authors and paired with GEU researchers while controlling for career stage, scholarly output and collaboration style. We constructed co-authorship networks for the CIFAR-GEU and Control groups and compared them over time. We also compared these two groups across several publication metrics to study the impact of stronger and longer-term networks. In this way, the control group acts as a proxy for changes in the field overall.

We hypothesized a denser co-authorship network in the CIFAR-GEU group, and this is certainly the case, especially looking at the increase in network density over time (Figure 2). Though both the CIFAR-GEU group and the Control group show increased collaboration over time (as part of a background trend for more collaborative research in general), the CIFAR-GEU network became much denser, much faster.

Looking at the groups pre-1987, it could be noted that researchers in the CIFAR-GEU group were already collaborating, while those in the control group were not. This is not surprising. When the program began in 1985, a few of the program founders were already in Canada and had worked together. It was the members of that core group who brought in others. In the control group however, there was no reason to expect collaboration because the Control group is composed of a very small subset of CIFAR co-authors who don't necessarily work together. For both groups, most co-authorships were established towards the middle of the study period (pre-2006). This is true for both the Control and CIFAR-GEU groups, but the effect is much stronger in the CIFAR-GEU group. Bringing researchers together and having them produce collaborative outputs is exactly what CIFAR strives to do. The networks show that CIFAR-GEU researchers are collaborating above and beyond the background growth in collaboration across the field.

We were also interested in the question of whether a denser network of collaboration strengthens the quality and reach of the research. The bibliometrics in Table 2 suggest that this is indeed the case. In terms of citation metrics, the CIFAR-GEU group was higher across all measures indicating that the CIFAR-GEU network produces high quality outputs that are cited more often than those produced by the Control group.

In terms of international collaboration, both the CIFAR-GEU and the Control groups collaborate internationally more than researchers in Canada in general. This is evidenced by a much higher percentage of publications with international collaboration (CIFAR-GEU 71.5% and Control 75.4%) than Canada as a whole (43%), which is in itself much higher than this metric for the United States (27%). This reflects a field of study where it is very difficult to publish in isolation of the global community. This is true more so today than when the CIFAR-GEU program started. At that time, the program founders proposed bringing in experts from outside Canada to bring together knowledge from the then disparate fields of physical cosmology, particle physics and gravitational theory. The idea was that no single lab or field contained all the skills necessary to tackle the biggest questions. In the early years, experts moved to Canadian universities from internationally renowned institutions such as Princeton,

Stanford, and Cambridge. When the top international astronomers and cosmologists joined Canadian institutions, it decreased the % of international collaborations. With this in mind, it is not surprising that the percentage of publications involving international collaboration was slightly higher in the Control group than the CIFAR-GEU group, especially in the early years of the program. However, when looking at the impact of international collaborations, the CIFAR-GEU group has a higher average number of citations for their internationally authored publications than the Control group.

The CIFAR-GEU group also had higher scores for metrics related to reach outside of academia. This was true for policy citations (policy documents that reference an output in the publication set of the group), blog mentions, news mentions and tweets. On the other hand, the Control group had a slightly higher % of Academic-Corporate collaboration, but as with International Collaboration, the impact of the Academic-Corporate collaborations that CIFAR researchers have, exceeds that of the control group. Finally, the Control group also had more patent citations (patents citing the scholarly output published by the group). SciVal gathers patent data from five of the largest patent offices: the European Patent Office (EPO), the US Patent Office (USPTO), the UK Intellectual Property Office (UK IPO), the Japan Patent Office (JPO) and the World Intellectual Property Organization (WIPO). However, none of these are Canadian and, as mentioned earlier, the CIFAR-GEU program was initially created to bring top astronomers and cosmologists to Canadian institutions. This made the whole group a little more Canadian and if these researchers filed patents only at the Canadian patent office, those patents weren't captured by SciVal.

The need for collaborative teams that bring together experts from different fields has increased as the complexity of the challenges researchers face has increased. Though there is mounting evidence of the benefits of collaborative research (e.g., Adler and Stewart, 2010; Beaver, 2004), it remains difficult to measure the impact of such teams, especially over long periods of time. Physicists have been working in teams to answer big questions for decades and, in this study, we endeavoured to measure the how effectively the CIFAR program Gravity & the Extreme Universe has created collaborations which resulted in high quality research outputs. We found stronger relationships created and maintained over time in the CIFAR-GEU group compared to the Control group. We also found that the outputs generated by the CIFAR-GEU group had on the whole, stronger metrics around citations, impact from international collaborations and reach beyond academia.

## Conclusions

The results of this study suggest that long term funding to convene the world's top researchers, produces stronger networks of collaboration and impact, even without massive investment in team grants that directly fund research. CIFAR's funding model focuses on convening program members and other national and international guests. Funding does not cover the costs of the research itself, but rather plays a complementary role within national research ecosystems. For example, in Canada federal funding from three discipline specific agencies largely supports the direct costs of research (salaries, equipment, etc.), while CIFAR supports the program meetings and seed funding for high-risk emerging areas (Innovation Science and Economic Development (ISED) Canada 2022).

The complex challenges facing researchers and society cannot be solved through disciplines where impact can be measured through publications, citations and bibliometrics alone. Indeed, the field of research impact is evolving rapidly and shifting away from traditional bibliometric indicators (Butler et al. 2017; Iping et al. 2022). In this study we chose to focus on a group whose goals for impact were largely focused on the academy and we used the tools available at the time to begin to measure reach outside of the academy. Applying this same approach with other programs that focus on problems rooted in societal issues remains an exciting next step. For example, many of CIFAR's newer programs have now been running for more than a decade and bring together multiple sectors and tackle challenges that reach well beyond the academy. It will be interesting to analyze their impacts, using emerging methods, and to contrast impact from this type of support with other funding models.

## Acknowledgements/Declarations

Deborah Buehler and Mark Daley previously worked at CIFAR. During the data collection phase of this project, Deborah Buehler had access to the CIFAR-GEU program's historical membership through her role as Research Information Manager. Similarly, Kyle Demes previously worked as a Research Consultant at Elsevier. At the time of writing, Deborah Buehler and Mark Daley were independent from CIFAR and Kyle Demes was independent from Elsevier. The authors did not receive financial support from any organization for the submitted work. The authors would like to thank Kate Geddie and Andrea Sharkey at CIFAR for helpful comments and advice.

## Author Contributions

All authors contributed to the study conception and design. Material preparation, data collection and analysis were performed by Deborah Buehler and Kyle Demes. The first draft of the manuscript was written by Deborah Buehler and all authors commented on previous versions of the manuscript. All authors read and approved the final manuscript.

TABLES

|  | Snowball* | Metric Definition |
|---|---|---|
| **Quality of Output** | | |
| Citation Count | Yes | Total citations received by the entity |
| Citations per Publication | Yes | Average citations received by an item of scholarly output |
| Cited Publications (%) |  | % of publications cited |
| Field-Weighted Citation Impact | Yes | Ratio of total citations received by the denominator output and total citations expected based on the average of the subject field. Thus, actual citation count relative to the expected world citation count |
| Output in Top 1% Citation Percentiles | Yes | Outputs that are amongst the top 1% most highly cited |
| **International Collaboration** | | |
| International Collaboration (%) | Yes | Percentage of outputs that have international co-authorship |
| International Collaboration Impact | Yes | Average citations received by the set of output that has international co-authorship. |
| **Reach beyond Academia** | | |
| Academic-Corporate Collaboration (%) | Yes | Percentage of outputs that have been co-authored researchers received by the outputs that have been co-authored by researchers from both academic and corporate or industrial, affiliations. |
| Academic-Corporate Collaboration Impact | Yes | Average citations received by the set of output that has academic-corporate co-authorship. |
| Citing-Patents Count |  | Count of patents citing the scholarly output published by the entity |
| Policy Citations |  | Number of policy documents that reference an output in the publication set of the entity. |
| Blog mentions |  | Number of times publications by the entity are mentioned in Blog posts. |
| News mentions |  | Number of times publications by the entity are mentioned in news articles. |
| Tweets |  | Number of times publications by the entity are mentioned in Tweets. |

*Table 1: Panel of bibliometric variables used to compare the CIFAR-GEU and Control groups and their publication sets. Snowball designation and definitions are from Elsevier (2016) and Colledge (2017)*

|  | **CIFAR-GEU** | Control |
|---|---|---|
| **Quality of Output** | | |
| Citation Count | **1,289,799** | 871,085 |
| Citations per Publication | **95.8** | 64 |
| Cited Publications (%) | **94.8** | 92.2 |
| Field-Weighted Citation Impact | **3.63** | 2.78 |
| Output in Top 1% Citation Percentiles | **1130** | 742 |
| | | |
| **International Collaboration** | | |
| International Collaboration (%) | 71.5 | **75.4** |
| International Collaboration Impact | **108.8** | 72.3 |
| | | |
| **Reach beyond Academia** | | |
| Academic-Corporate Collaboration (%) | 4.2 | **5.9** |
| Academic-Corporate Collaboration Impact | **224.2** | 130.8 |
| Citing-Patents Count | 191 | **1074** |
| Policy Citations | **99** | 88 |
| Blog mentions | **3,019** | 1,217 |
| News mentions | **5,648** | 3,739 |
| Tweets | **53,195** | 31,490 |

*Table 2: Bibliometrics for CIFAR-GEU and Control group publication sets from 1996 to 2021.*

FIGURES

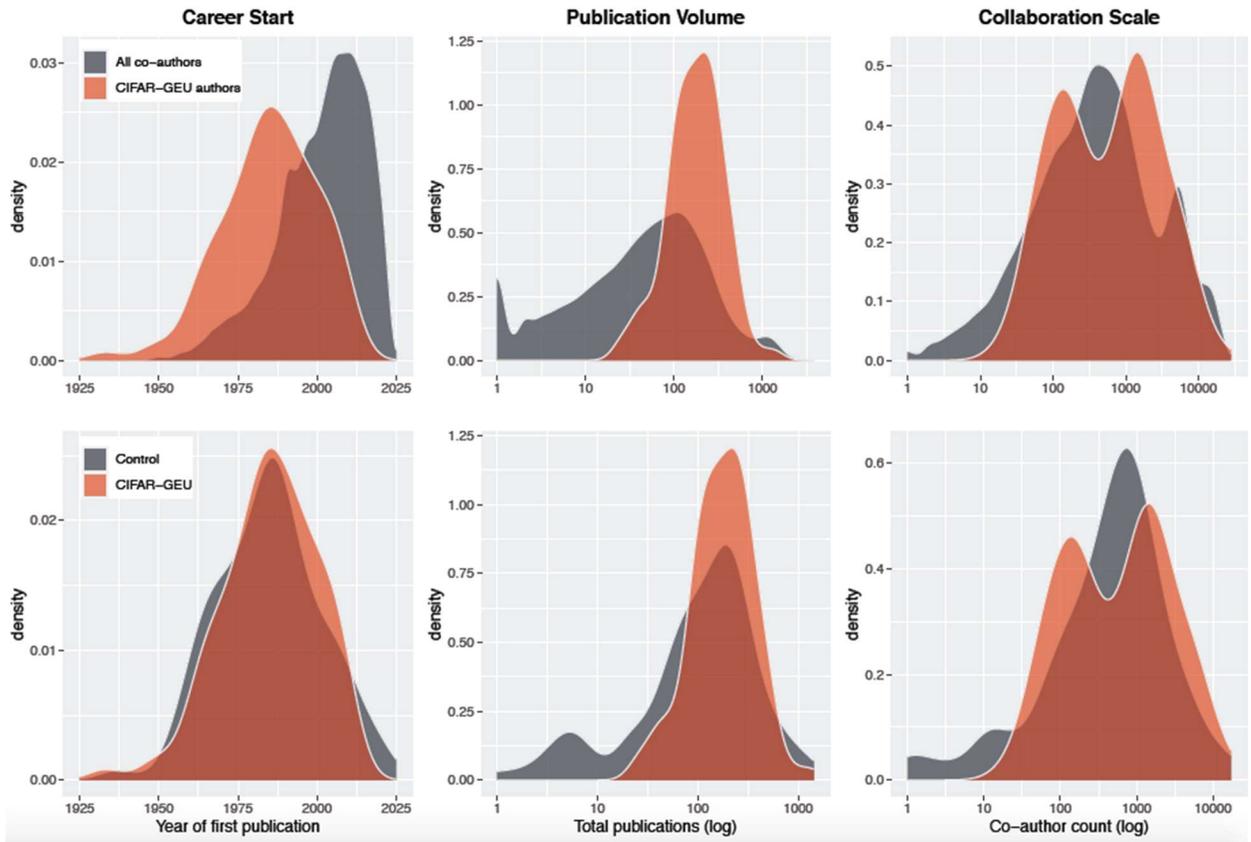

*Figure 1: Density plots comparing Scopus author-level metrics for CIFAR-GEU researchers (red, n=93) with the entire pool of co-authors (grey, N>26,177).*

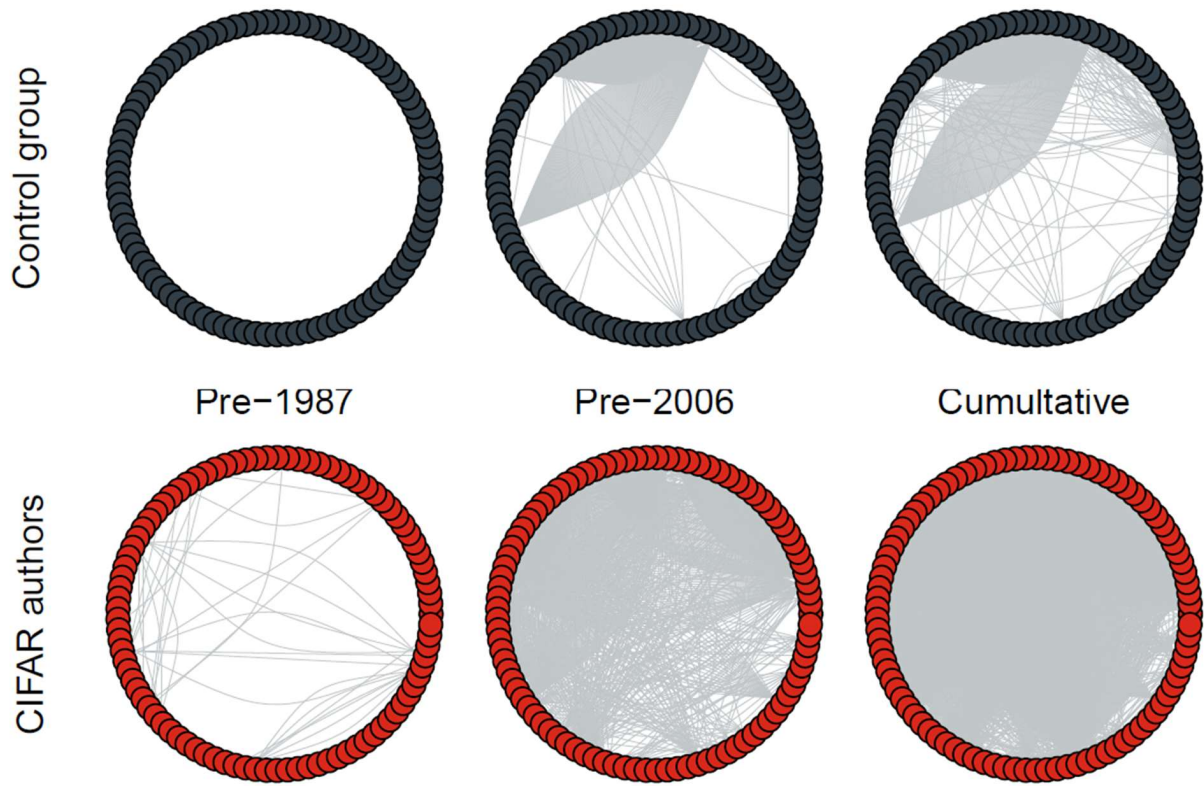

*Figure 2:* CIFAR-GEU and Control group co-author networks built from papers published just before the program was started on the left, the per-2005 in the middle and the cumulative publications to 2022 on the right.

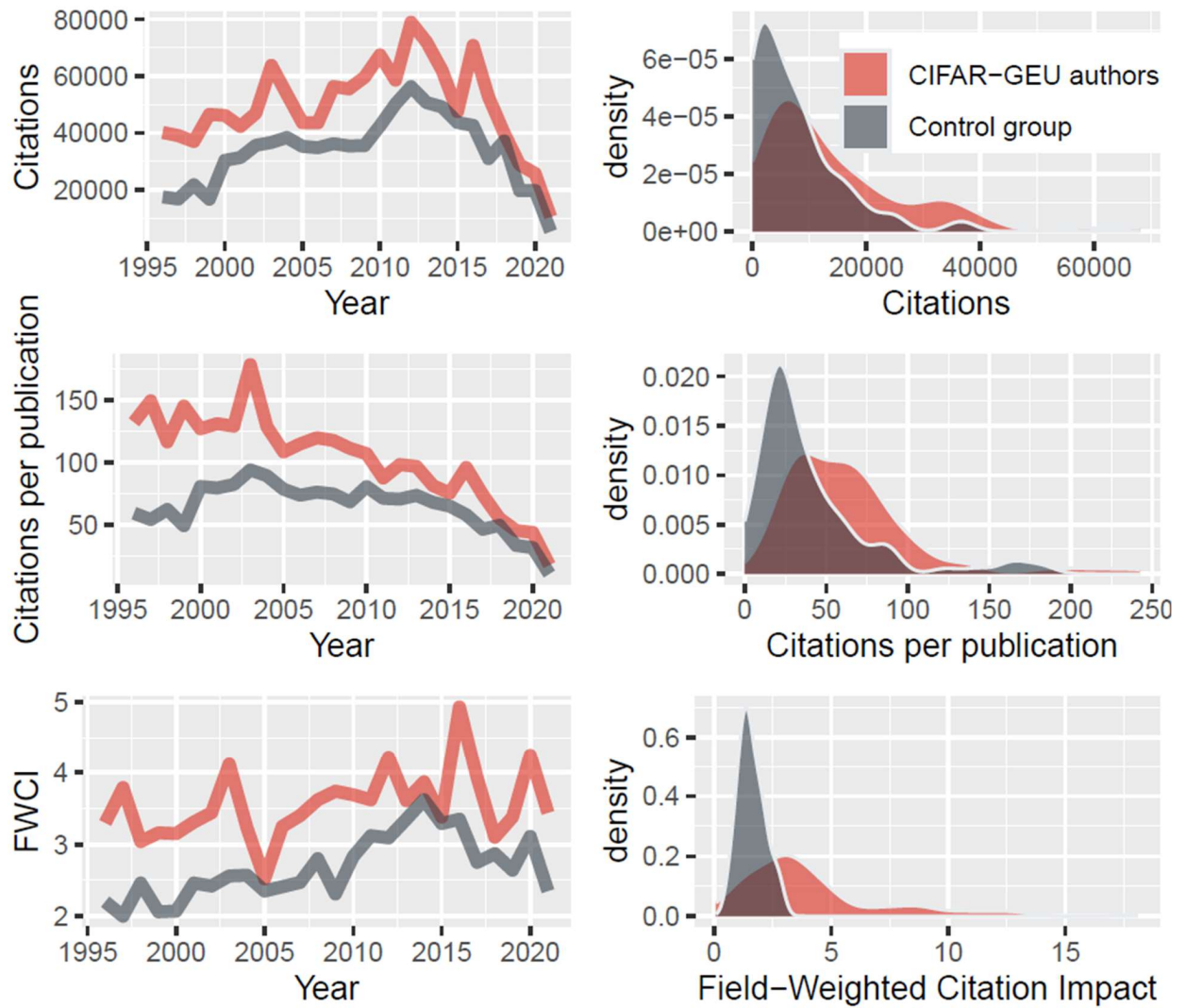

*Figure 3: Left panel shows group-level data for the citation metric over time (i.e., each unique publication is only counted once to generate a group-level citation metric) and the right panel shows the distribution of author-level metrics for the same bibliometric indicators (i.e., all of an author's publications are used to calculate their metrics). Data span from 1996 (first year available in SciVal) to 2021 (last 'complete year' available in SciVal at time of analysis). In both figures CIFAR-GEU is denoted with red fill, while the Control group is denoted with grey fill.*